\documentclass[prb,
superscriptaddress,showpacs,amsmath,amssymb]{revtex4}

\usepackage{graphicx}

\begin{document}

\title{Topological Lifshitz transitions}

\author{G.E.~Volovik}
\affiliation{Low Temperature Laboratory, Aalto University,  P.O. Box 15100, FI-00076 Aalto, Finland}
\affiliation{Landau Institute for Theoretical Physics, acad. Semyonov av., 1a, 142432,
Chernogolovka, Russia}

\date{\today}

\begin{abstract}
{ Different types of Lifshitz transitions are governed by topology
in momentum space. They involve the topological transitions 
with the change of topology of Fermi
surfaces, Weyl and Dirac points, nodal ines, and also the
transitions between the fully gapped states.
}
\end{abstract}

\maketitle

\section{Introduction}

\begin{figure}
\centerline{\includegraphics[width=0.8\linewidth]{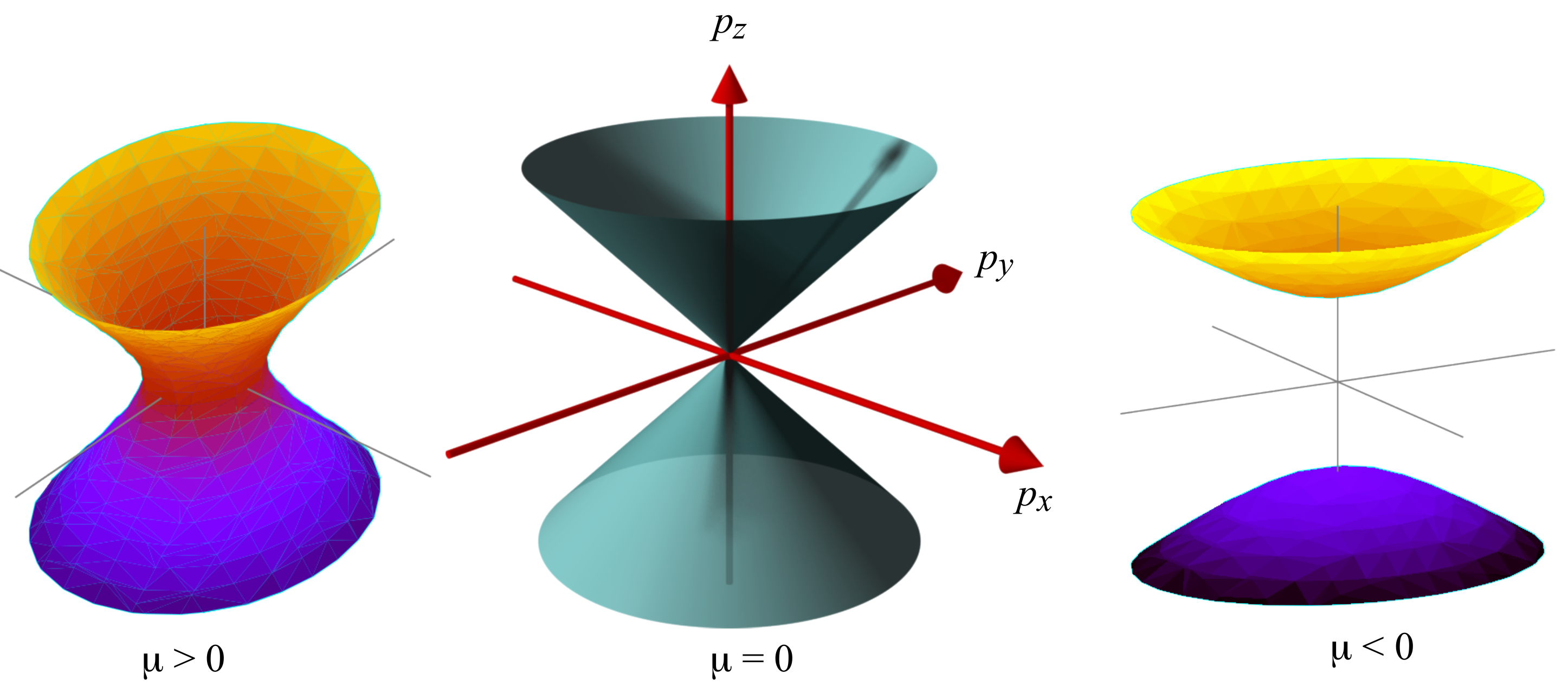}}
\medskip
\caption{Example of the topological transition with the change of the Fermi surface topology -- disruption of the "neck" of
a Fermi surface -- presented in the Lifshitz paper.\cite{ILifshitz1960}
}
\label{ParHypLifshitzFig}
\end{figure}

Originally the Lifshitz transitions were related to the electron transitions at $T=0$ in metals, at which the topology of the Fermi surface of the metal changes abruptly. Fig. \ref{ParHypLifshitzFig} demonstrates an example, when
during the continuous change of the parameter of the system -- the chemical potential $\mu$ in Eq.(\ref{ExamplleLifshitzTransition}) 
\begin{equation}
\epsilon_{\bf p}  = \frac{p_x^2 + p_y^2 -p_z^2}{2m}-\mu   
\,,
\label{ExamplleLifshitzTransition}
\end{equation}
 the  disruption of the Fermi surface occurs at $\mu=0$.\cite{ILifshitz1960} Such discontinuity at Lifshitz transitions gives rise to anomalies in the electron characteristics
of  metals. In the other electronic materials, such the topological semimetals, topological insulators and topological superconductors, different types of the Lifshitz transitions take place.\cite{Volovik2007} They involve 
the other types of zeroes in the energy spectrum in addition to or instead of the Fermi surface, such as 
flat bands, Weyl and Dirac point nodes, Dirac nodal lines, zeroes in the spectrum of edge states, Majorana modes, etc. 

Let us start with the Lifshitz transitions, which involve the dispersional (flat) band, where all the states within some region of the $3D$ or  $2D$  momenta have zero energy.  The flat bands may emerge in bulk due to interaction between electrons\cite{Khodel1990,Volovik1991,Nozieres92} or on the surface of materials due to nontrivial topology of the electronic spectrum in bulk.\cite{Ryu2002,SchnyderRyu2011,HeikkilaVolovik2011,HeikkilaKopninVolovik2011,SchnyderBrydon2015} 

\section{Flat bands near the conventional Lifshitz transition}

We illustrate the formation of the flat band solution due to electron-electron 
interaction using the Landau energy functional in terms of the distribution function, 
$ {\cal E}\{n({\bf p})\}$. The variation of the functional gives the equation for $n_{\bf p}$ and the quasiparticle energy $\epsilon_{\bf p}$:
\begin{equation}
\delta {\cal E}\{n({\bf p})\} = \int \frac{d^d p}{(2\pi \hbar)^d}~ \epsilon_{\bf p} \delta n_{\bf p} =0
\,.
\label{variation}
\end{equation}
In general, since the quasiparticle distribution function is constrained by the
Pauli principle, $0\leq n_{\bf p}\leq 1$, there are two classes of
solutions of the variational problem. One class is $\epsilon_{\bf
  p}=0$, which in conventional metals determines the Fermi surfaces; and another one is $\delta
n_{\bf p}=0$, where either $n_{\bf p} =0$ or $n_{\bf p} =1$. The regions  with  $n_{\bf p} =0$ and $n_{\bf p} =1$ are separated by the Fermi surface. 

\begin{figure}
\centerline{\includegraphics[width=0.5\linewidth]{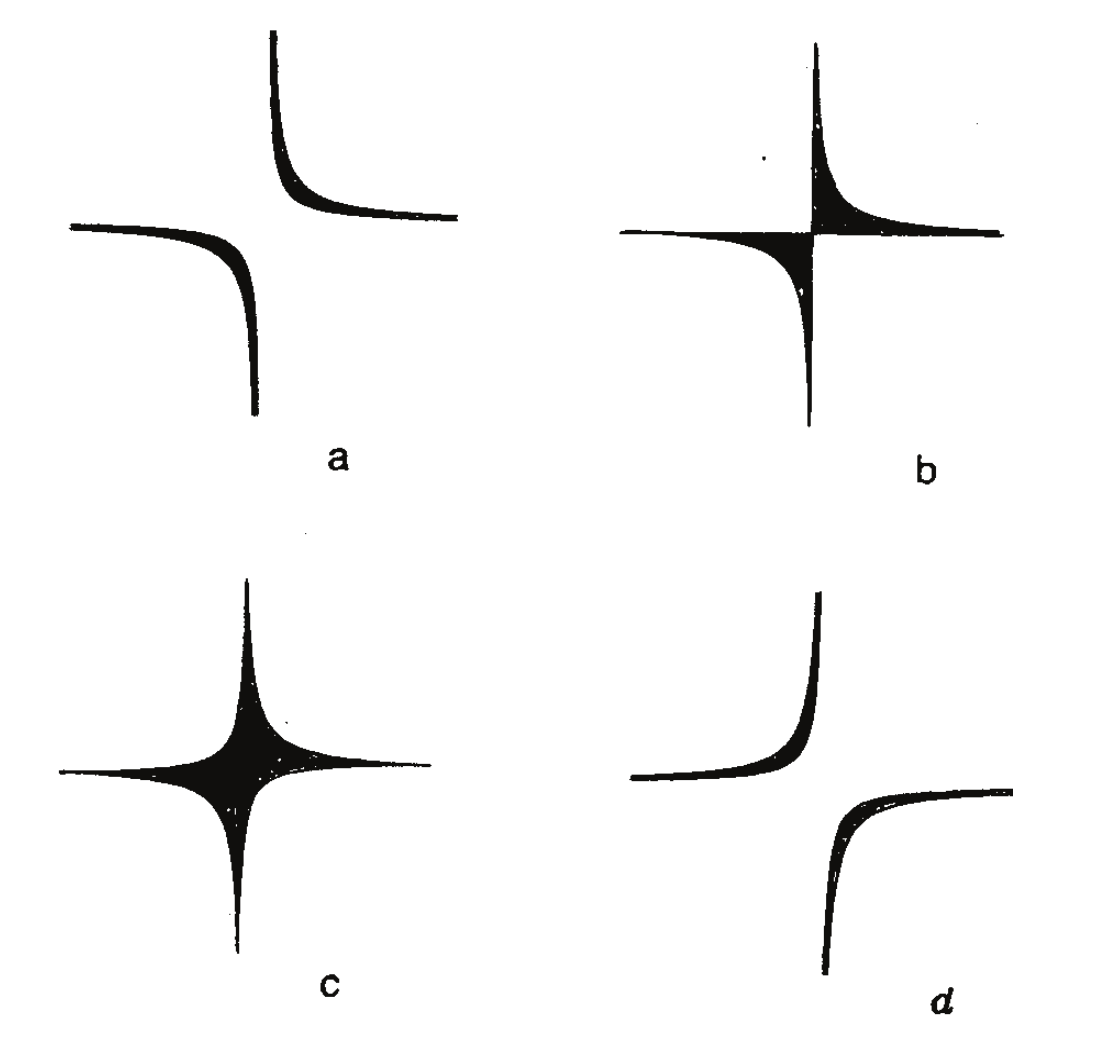}}
\medskip
\caption{Series of Lifshitz transitions with formation and evolution of the flat bands near the saddle point 
of 2D spectrum, when the chemical potential $\mu$ changes approaching and crossing the Lifshitz transtion point in the unperturbed spectrum. When $\mu$ is close enough to the saddle point, the quantum phase transition(s) occur, when two flat bands are formed simultaneously (a), or one after the other. With further change of the chemical potentials these two flat bands touch each other in a new Lifshitz transion (b) and form single flat band (c). The latter splits in another Lifshitz  transition (d). Fnally, when $\mu$ is far enough from the saddle point, these two flat bands disappear sumultaneously or one by one in the other Lifshitz transition(s). Altogether in the vicinity of the Lifshitz transition in the noninteracting system, from four to six Lifshitz transitions are expected, which involve the flat bands.}
\label{FCFig}
\end{figure}

When the interaction between electrons is strong enough, one may expect the
novel Lifshitz  transition at which the solution
$\epsilon_{\bf p}=0$ starts to spread from the Fermi surface to a finite region in momentum space, i.e. the Fermi surface trnsforms to
the flat band. The flat band solution can be illustrated using simple functional describing the  contact interaction of electrons:
\begin{equation}
 {\cal E}\{n({\bf p})\} = \sum_{\bf p}\left[\epsilon^{(0)}_{\bf p}  n_{\bf p}  + 
\frac{1}{2} U \left(n_{\bf p}-\frac{1}{2}  \right)^2\right]
\,.
\label{Saddle}
\end{equation}
It is argued that the flat band is more easily formed, when the non-perturbed spectrum
(i.e. at $U=0$) is close to the  Lifshitz transition 
discussed in previous section.
\cite{Volovik1994,Yudin2014}
Here we consider the 2D case, when the non-interacting spectrum, $\epsilon^{(0)}_{\bf p}=\frac{p_x p_y}{m}
- \mu$, experiences the conventional Lifshitz transion at $\mu=0$, while the electron-electron interaction may lead to multiple  Lifshitz transitions.

\begin{figure}
\centerline{\includegraphics[width=0.7\linewidth]{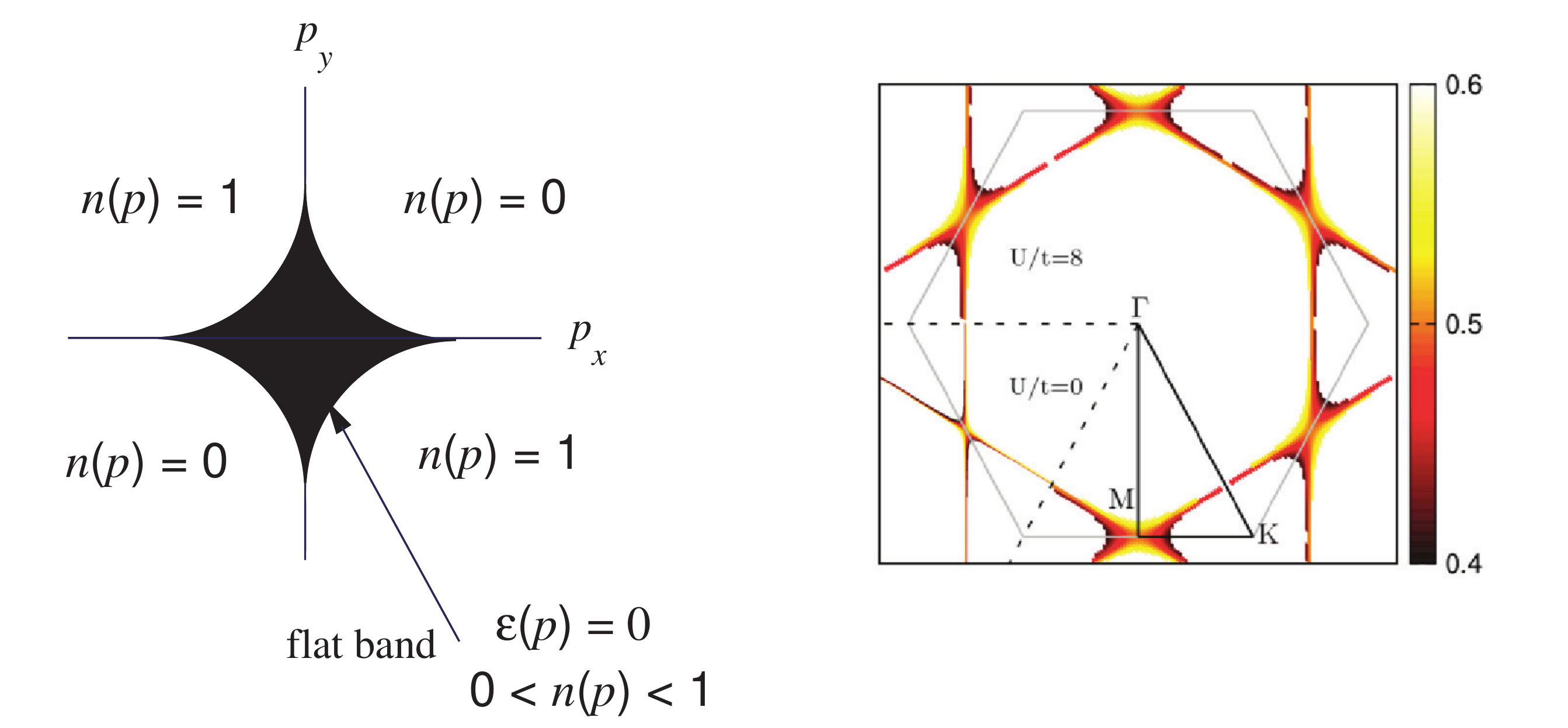}}
\medskip
\caption{Flat band emerging near a saddle point. ({\it Left}): from
  the simplified Landau-type theory in
  Eqs. (\ref{Saddle})-(\ref{VariationSolution}) with $\mu=0$. The
  flat band is concentrated in the black region.  ({\it Right}): from
  the numerical solution of the Hubbard model,\cite{Yudin2014}
  showing the spectral function within the reciprocal space of an interacting triangular lattice.
The lower left sextant corresponds to the noninteracting case $U=0$. For large $U$ the band flattening
is clearly seen near the saddle points. }
\label{SaddleFig}
\end{figure}

The functional (\ref{Saddle}) has always a flat band solution with $0< n_{\bf p} <1$:
\begin{equation}
 \epsilon_{\bf p}=\frac{\delta{\cal E}}{\delta n_{\bf p}} =\epsilon^{(0)}_{\bf p}   +U  \left(n_{\bf p}-\frac{1}{2}  \right)=0 \,\,,   \,\, n_{\bf p}  = \frac{1}{2} - \frac{\epsilon^{(0)}_{\bf p}}{U}~~,~~ 0< n_{\bf p} <1 \,.
\label{VariationSolution}
\end{equation}
For $|\mu| > U/2$, there are two isolated flat bands. They merge at the Lifshitz transition, which   takes place at  $\mu=- U/2$, and split again at $\mu= U/2$, see Fig. \ref{FCFig}.
In the more realistic situation, there should be also the topological transitions, at which each of the two isolated flat bands emerge from the Fermi surfaces and disappear, resulting altogether to 6 Lifshitz transitions involving the flat bands. The formation of the flat band in the simplified Landau model  is supported by  numerical simulations of  the Hubbard model,\cite{Yudin2014}  see Fig. \ref{SaddleFig}, where the model  flat band at $\mu=0$ in Fig. \ref{SaddleFig} ({\it left}) is compared with the result of the numerical simulations of  the Hubbard model in the vicinity of the van Hove singulrity in Fig. \ref{SaddleFig} ({\it right}).\cite{Yudin2014} 

Since the flat band has singular density of electronic states, one may expect the enhancement of the
superconducting transition temperature in the materials with flat band.\cite{Khodel1990,HeikkilaKopninVolovik2011}
The reason for that is that instead of the exponentially suppressed gap formed in the material with the Fermi surface: 
\begin{equation}
\Delta = E_0  \exp\left(- \frac{1}{ g N_F}\right) \,,
\label{exponent}
\end{equation}
in the system with the flat band the gap is linearly proportional to the interaction
 strength $g$:
\begin{equation}
 \Delta = \frac{ gV_d}{2(2\pi \hbar)^d}\,.
\label{linear}
\end{equation}
Here $N_F$ is the density of states in normal metal; $d$ is the dimension of the metal; and $V_d$ is the volume of the flat band.

It is remarkable that the 203 K superconductivity in H$_3$S has been reported at high pressure\cite{Drozdov2015}, where the Fermi surface is close to the van 
Hove singularity.\cite{Pickett2015,Bianconi2015} 
This suggests that the  enhancement of the transition temperature in H$_3$S can be due to the flat band formed in the vicinity of the singularity. It is not excluded that the flat band is responsible for superconductivity in FeSe,\cite{FeSe}  and for the traces of the high-$T_c$ superconductivity in graphite.\cite{EsquinaziHeikkila2014,Esquinazi2013a,Esquinazi2013b,Esquinazi2013c} 

In this section we considered the additional topological transitions, which emerge due to extension of the Fermi surface to the manifold of zeroes of higher dimension.  The other new types of Lifshitz transitions takes place when the zeroes of the lower dimensionalities are involved, such as the topologically protected nodal lines and point nodes. Let us consider as an example the $d$-wave superconductivity in the system close to the conventional Lifshitz transition.

\section{ Lifshitz transition from gapped state to state with Weyl points}
\label{GappedToWeylSec}

\begin{figure}
\centerline{\includegraphics[width=0.8\linewidth]{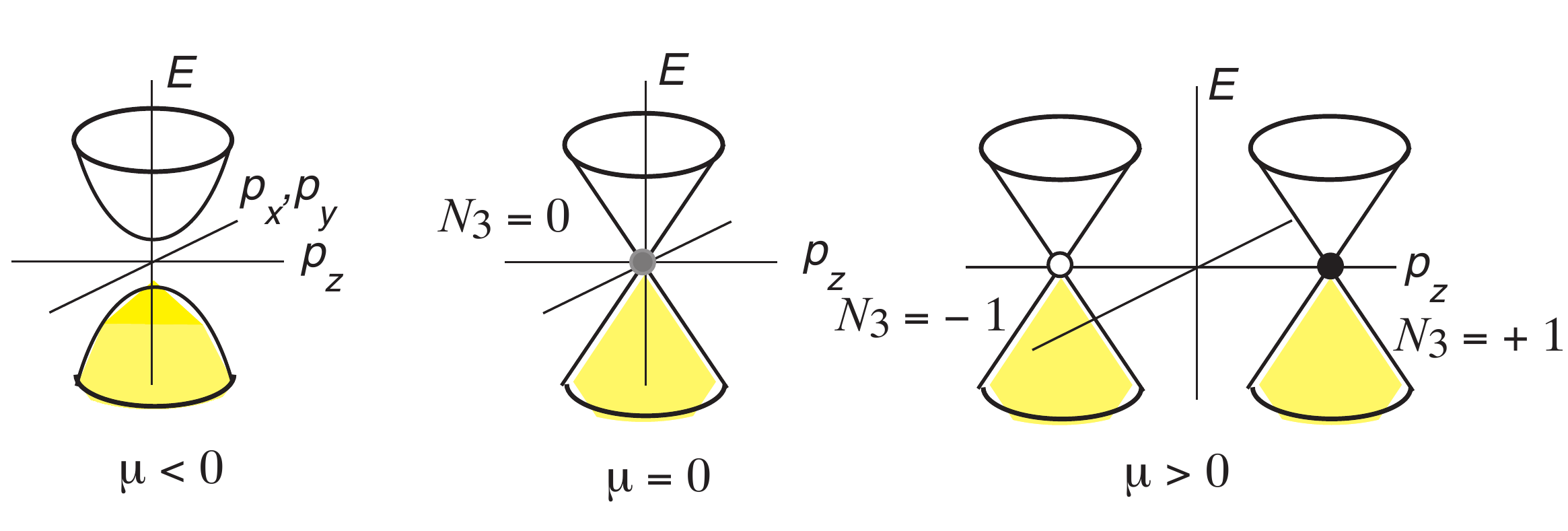}}
\medskip
\caption{Lifshitz transition at $\mu=0$ between a fully-gapped vacuum and a
vacuum with topologically protected Weyl points. At $\mu=0$, there appears
a marginal Fermi point with topological charge $N_3=0$. For $\mu>0$, the
marginal Fermi point splits into two Weyl points characterized by nonzero topological
invariants $N_3=\pm 1$. 
}
\label{FromDiracToWeyl}
\end{figure}

Let us consider two examples of the new types of Lifshitz transitions, which may occur in the $p$-wave and $d$-wave superconductors. The instructive model Hamiltonian for spinless fermions with the chiral $p$-wave order parameter is
\begin{equation}
H= \tau_3 \epsilon_{\bf p} + c(p_x \tau_1 + p_y\tau_2) \,\, \,\,, \,\,  \,\, \epsilon_{\bf p}  = \frac{p_x^2 + p_y^2 +p_z^2}{2m}-\mu   
\,,
\label{Pwave}
\end{equation}
where $\tau_a$ are Pauli matrices in the Bogoliubov-Nambu particle-hole space.

In the normal state, where the "speed of light" parameter $c=0$, there is the conventional Lifshitz transition at $\mu=0$, with formation of the Fermi surface at $\mu>0$. In the superconducting state, at $\mu=0$ one has the new type of Lifshitz transition: it separates the massive Dirac state at $\mu<0$ and the state with two Weyl points at $\mu>0$, Fig.\ref{FromDiracToWeyl}. The Weyl points  are at  ${\bf p}_\pm=(0,0,\pm \sqrt{2m\mu})$. They are topologically protected, since they correspond to the Berry phase monopole in momentum space.\cite{Volovik1987,Volovik2003}
 
\begin{figure}
\centerline{\includegraphics[width=0.5\linewidth]{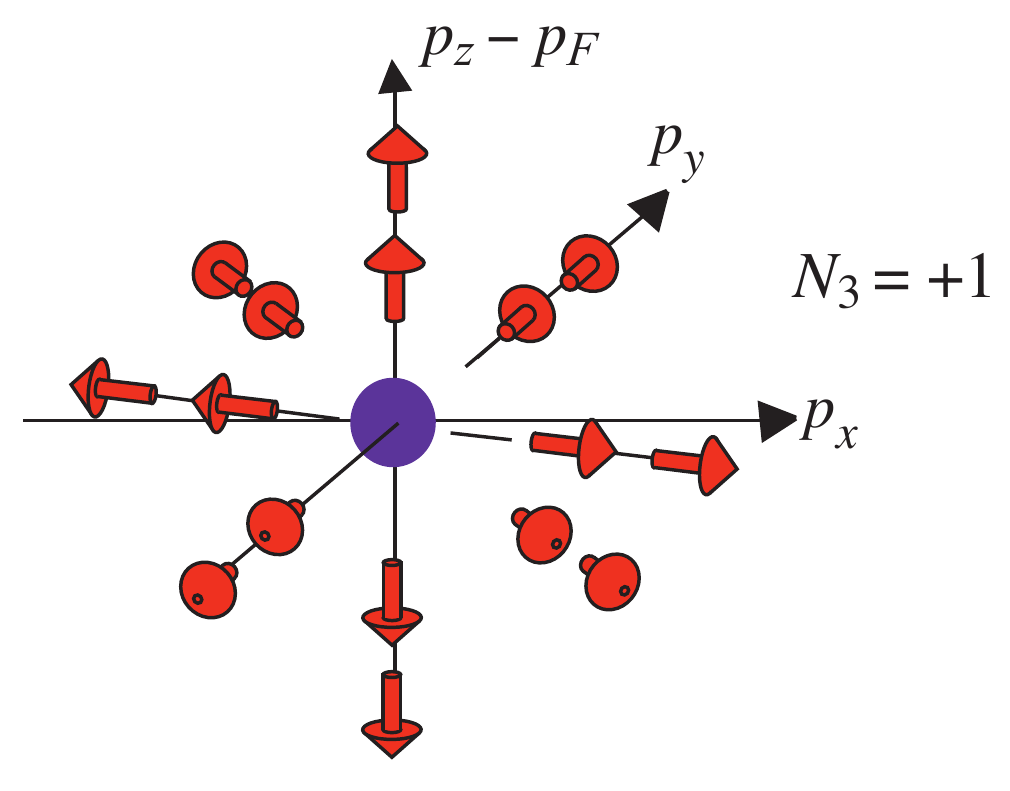}}
\medskip
\caption{Illustration of the topological stability of the Weyl point. Arrows show the direction of the vector field 
${\bf m}({\bf p})$ around the
Weyl point at ${\bf p}=p_F \hat{\bf z}$. The structure of the field froms  represents the hedgehog with the spines of the hedgehog directed outward. The degree of mapping of the 2D surface $S_2$ to the $2D$ sphere of unit vector $\hat{\bf m}$ represents the topological invariant $N_3$ in Eq.(\ref{NeutrinoInvariant}). For this Weyl point $N_3=1$.
The Weyl point at ${\bf p}=-p_F \hat{\bf z}$ corresponds to the hedgehog with spines inward and the opposite topological invariant, $N_3=-1$.}
\label{FP1Fig}
\end{figure}

To illustrate the topological stability of the Weyl points, let us rewrite the $2\times 2$ Hamiltonian (\ref{Pwave}) in the general form
\begin{equation}
H=
\tau^i m_i({\bf p})\,\, , \,\, i=(1,2,3) \,,
\label{NeutrinoGeneral}
\end{equation}
where in our case the components of the vector ${\bf m}$ are
\begin{equation}
m_1=cp_x \,\,, \,\, m_2=cp_y \,\,, \,\,  m_3=\frac{p^2}{2m}-\mu\,.
\label{SpinlessFermionsHamiltonianVector}
\end{equation}
The Weyl points ${\bf p}_\pm=(0,0,\pm \sqrt{2m\mu})$ are singular points in momentum space -- the centers of the hedgehogs in the vector field ${\bf m}({\bf p})$, see Fig.~\ref{FP1Fig}.
In the center of the hedgehog the direction of ${\bf m}$ is not defined, thus  ${\bf m}({\bf p}_\pm)=0$ and the  energy of quesiparticles at these points in momentum space should be zero. Close to zero energy the spectrum is similar to that of the massless relativistic Weyl fermions, this is the reason why these points are called the Weyl points. The topological invariant which
protects the Weyl points from formation of the gap  is the winding number of the mapping of the sphere
$S_2$ around the singular point in
${\bf p}$-space to the 2-sphere of the unit vector $\hat {\bf m}={\bf
m}/|{\bf m}|$:
\begin{equation}
N_3=\frac{1}{8\pi}e_{ijk}\int_{S_2} dS^k ~\hat{\bf
m}\cdot \left({\partial \hat{\bf m}\over\partial {p_i}} \times {\partial
\hat{\bf m}\over\partial {p_j}}\right)~.
\label{NeutrinoInvariant}
\end{equation}
One can check that the topological charges of the two Weyl points are
$N_3=\pm 1$.

 In relativistic physics, the Weyl point with $N_3=1$ describes the right-handed
chiral particles (quarks and leptons) obeying the Hamiltonian  $H=c {\mbox{\boldmath$\sigma$}} \cdot {\bf p}$, i.e. ${\bf m}({\bf p})=c{\bf p}$ where $c$ is the speed of light. The Hamiltonian  $H= -c{\mbox{\boldmath$\sigma$}} \cdot {\bf p}$,  i.e. ${\bf m}({\bf p})=-c{\bf p}$, describes the  left-handed
chiral particles which have the opposite topological charge, $N_3=-1$. 
The discussed Lifshitz transition in relativistic system has been considered in Ref.\cite{KlinkhamerVolovik2005a}.

\section{Lifshitz transition between Dirac line and Weyl points}

\begin{figure}
\centerline{\includegraphics[width=0.8\linewidth]{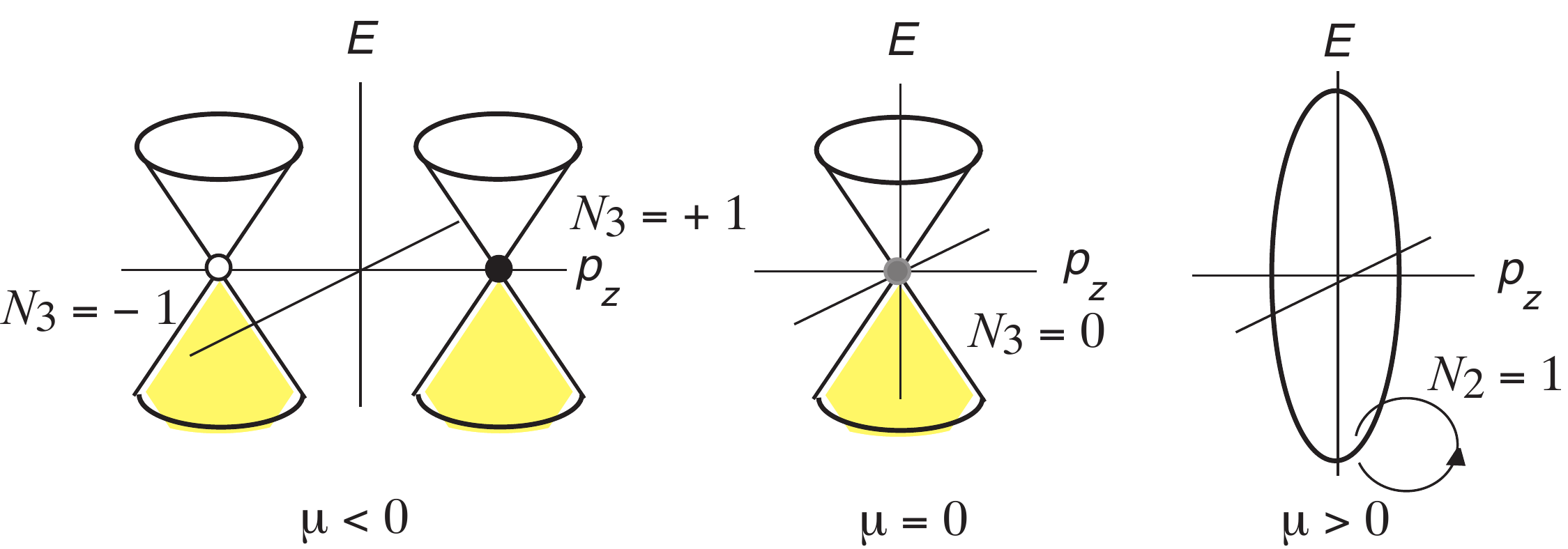}}
\medskip
\caption{Lifshitz transition at $\mu=0$ from the state with two Weyl points ({\it left}) to the state with Dirac nodal line ({\it right}) in $d_{zx} +i d_{zy}$ superconductor.
}
\label{FromWeylToLine}
\end{figure}

Another type of Lifshits transition takes place in case of $d_{zx} +i d_{zy}$ superconductivity, if it emerges in the vicinity of the saddle point Lifshitz transition in Fig. \ref{ParHypLifshitzFig}. The model Hamiltonian for this case is:
\begin{equation}
H= \tau_3 \epsilon_{\bf p} + cp_z(p_x \tau_1 + p_y\tau_2) \,\, \,\,, \,\,  \,\, \epsilon_{\bf p}  = \frac{p_x^2 + p_y^2 -p_z^2}{2m}-\mu   
\,,
\label{SaddleSC}
\end{equation}
In the normal state the conventional Lifshitz transition with the reconstruction of the  Fermi surface occurs at $\mu=0$.
In the superconducting state, the topological Lifshitz transition at $\mu=0$ separates the state with two Weyl points at $\mu<0$ from the state with the Dirac nodal line at $\mu>0$, Fig. \ref{FromWeylToLine}.
The topologically protected Weyl points are at  ${\bf p}_\pm=(0,0,\pm \sqrt{-2m\mu})$, while the Dirac line is at $p_z=0$, $p_x^2 + p_y^2 = 2m\mu$.  The Dirac line is also topologically protected, though by different topological invariant.
To see that let us consider one of the planes, which crosses the Dirac line, for example, the plane  $p_y=0$. This plane contains two Dirac points -- the singular points of the Namiltonian
\begin{equation}
H= \tau_3 \left(\frac{p_x^2 -p_z^2}{2m}-\mu\right) + cp_zp_x \tau_1 \,.
\label{SaddleSCprojection}
\end{equation}
The singular points are at $(p_x,p_z)=(\pm \sqrt{2m\mu},0)$.
The topological invariant for each of these two points can be written as the contour integral around the singular point:
\begin{equation}
N_2=  {\bf tr} \oint_C \frac{\mathbf dl}{4\pi i} \cdot[\tau_2 H^{-1}({\bf p}) \partial_l H({\bf p})],
\label{eq:N2}
\end{equation}
For these two points one has $N_2=\pm 1$, which supports the stability of the nodes. Since the choice of the plane is arbitrary, the topological stability is  valid  for any point on the Dirac line. The Dirac line is described by the invariant $N_2=1$ in Eq.(\ref{eq:N2}), where the contour $C$ is around the nodal line and instead of $\tau_2$ matrix there is the proper combination of the $\tau_1$ and $\tau_2$ matrices.

Similar transition has been discussed for the model of topological semimetal.\cite{LimMoessner2016}

\section{Lifshitz transition through Dirac conical points   with quadratic touching}
\label{QuadraticTouching}

Another type of Lifshitz transitions can be studied on example of the bilayer graphene described by the following
$2D$ Hamiltonian:\cite{Manes2007}
\begin{equation}
H= \tau_1 (p_x^2 -p_y^2+ vp_x) +\tau_2(2p_xp_y -vp_y)\,.
\label{bilayer}
\end{equation}

\begin{figure}
\centerline{\includegraphics[width=1.0\linewidth]{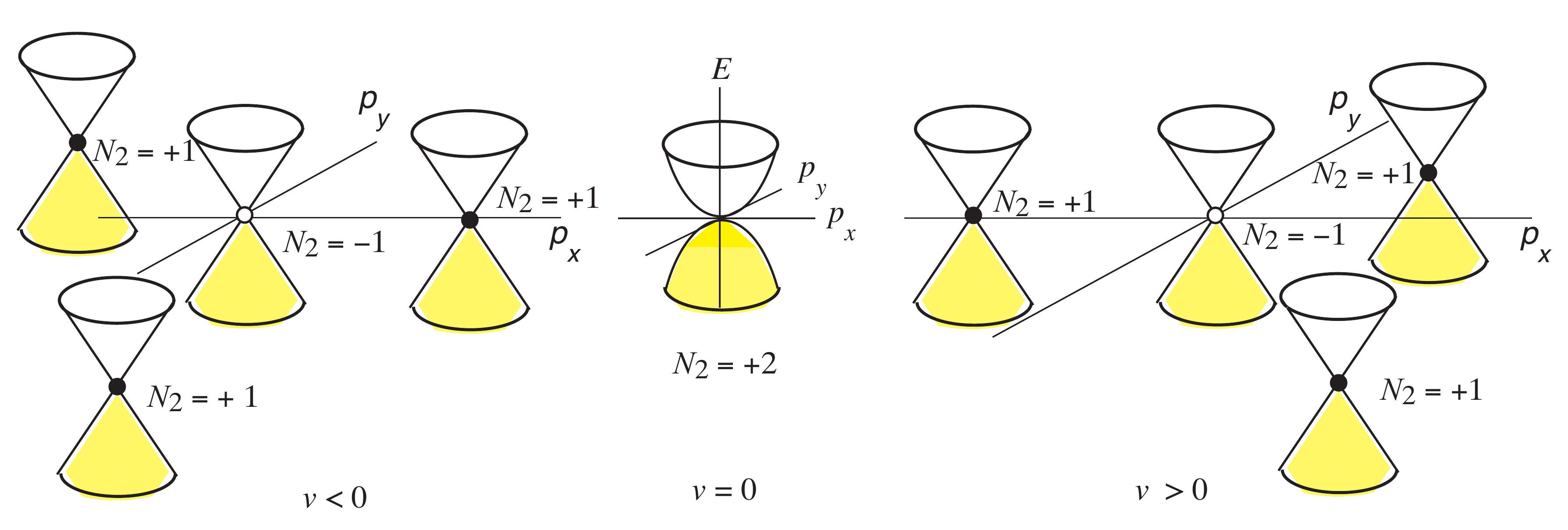}}
\medskip
\caption{The Lifshitz transition at which 4 Dirac points with topological charges $N_2=\pm 1$ merge together and then split again. At the point of Lifshitz transition the Dirac spectrum with quadratic touchning is formed, which is characterized by  topological charge $N_2=+2$.
}
\label{TrigonalFig}
\end{figure}

At $v\neq 0$ the spectrum contains four Dirac points described by topological charge $N_2$ in Eq.(\ref{eq:N2}), where the $\tau_2$ matrix is substituted by the $\tau_3$ matrix. One point with $N_2=-1$ is at the center (which corresponds to the $K$-point of graphene), while three others with  $N_2=+1$ each
are away from the $K$ point. This is the so-called trigonal warping. The total topological charge of all four Dirac points is $N_2=+1+1+1-1=+2$. When the parameter $v$ crosses zero, the spectrum is reflected. At the point of Lifshitz transition, i.e. at $v=0$, the spectrum contains the point node with multiple topological charge $N_2=+2$. 
It corresponds to the 2D Dirac fermions with quadratic touching,
$E_\pm=\pm p^2$, see Sec. 12.4.3 in Ref. \cite{Volovik2003}, while in general the  multiple topological charge $|N_2| > 1$ gives rise to Dirac fermions with the higher order touching point, $E^2 \sim p^{2|N_2|}$.\cite{HeikkilaVolovik2011} 
The similar behavior of the spectrum takes place also in the 3D case,\cite{VolovikKonyshev1987,Volovik2003} where the Weyl point with multiple topological charge $|N_3| >1$ has the spectrum $E^2=c^2p_\parallel^2 + \gamma p_{\perp}^{2|N_3|}$.

\begin{figure}
\centerline{\includegraphics[width=0.5\linewidth]{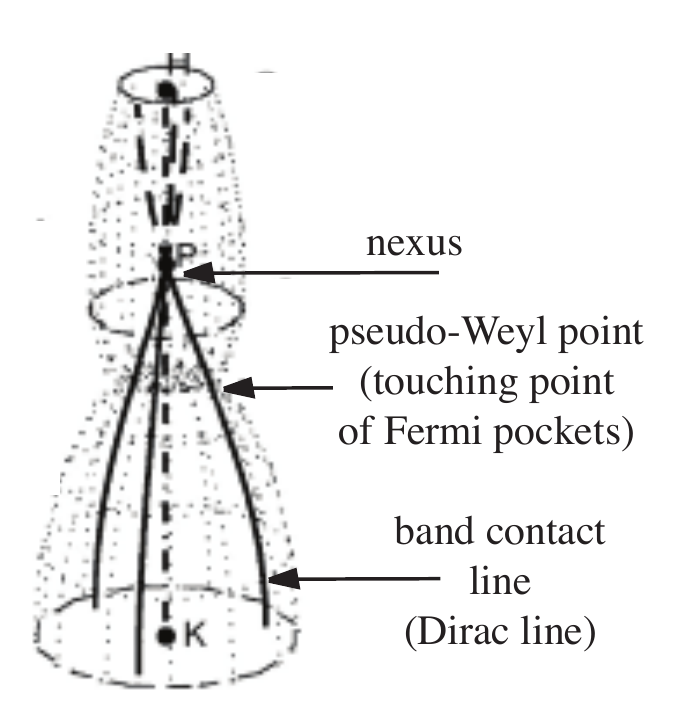}}
\medskip
\caption{
Nexus - the point in momentum space where four Dirac lines -- the lines of contacts of two bands characterized by the topological charge $N_2$ -- merge together 
(from Ref.\cite{Mikitik2006}). There is the one to one correspondence between the nexus in Bernal graphite and the Lifshitz transition in bilayer graphene in Fig. \ref{TrigonalFig}: the role of the parameter $v$ in $2D$ system is played fo the $p_z$ component of the momentum in the $3D$ system. This figure also shows the
points on Dirac lines, where the Fermi surface pockets touch each other forming the conical Fermi surface. These conical points are topologically protected by invariant $N_2$ (or by the invariant $N_3({\rm pseudo})$ in Eq.(\ref{NeutrinoInvariantPseudo})). Such pseudo-Weyl points appear in different cases including 
the Rashba model in Fig. \ref{RashbaMikitikFig}.
}
\label{GraphiteFig}
\end{figure}

In the bilayer graphene another interesting Lifshitz transition can be realized, where the intermediate state is a monkey saddle with dispersion $p_x^3 - 3p_xp_y^2$: at this transition three van Hove saddles merge
forming an elliptical umbilic elementary catastrophe.\cite{Shtyk2016}

In the 3D materials the role of the parameter of the Lifshitz transition can be played by the $p_z$ component. Then one has the correspondence between the Lifshitz transition in the $2D$ materials discussed in this Section and the singular point in the spectrum of the $3D$ material. In case of Lifshitz transition in Fig. \ref{TrigonalFig}, the corresponding singularity in the $3D$ spectrum is the so-called nexus in the Bernal graphite. The  Bernal graphite can be considered as periodic stacking of the graphene bilayers. Then the parameter $v$ becomes dependent on $p_z$, and  at some value of $p_z$ the function $v(p_z)$ changes sign. This is the nexus point, where four Dirac nodal lines merge,\cite{Mikitik2006,Mikitik2008,Mikitik2014,HeikkilaVolovik2015a} see Fig. \ref{GraphiteFig} (note that due to extra matrix elements the Dirac lines are shifted from the zero energy levels
and become situated inside the electron or hole pockets, but they remain to be
the band contact lines,\cite{Mikitik2006,Mikitik2008} which are described by the properly determined topological charge $N_2=\pm 1$, or in the other language by the 
 Berry phase $\Phi=\pm \pi$). 
 This is one of many examples, 
 when the  Lifshitz transition is equivalent to the topological defect in the higher dimensional momentum space. 

\section{Lifshitz transition in Rashba model and pseudo-Weyl fermions}

\begin{figure}
\centerline{\includegraphics[width=0.7\linewidth]{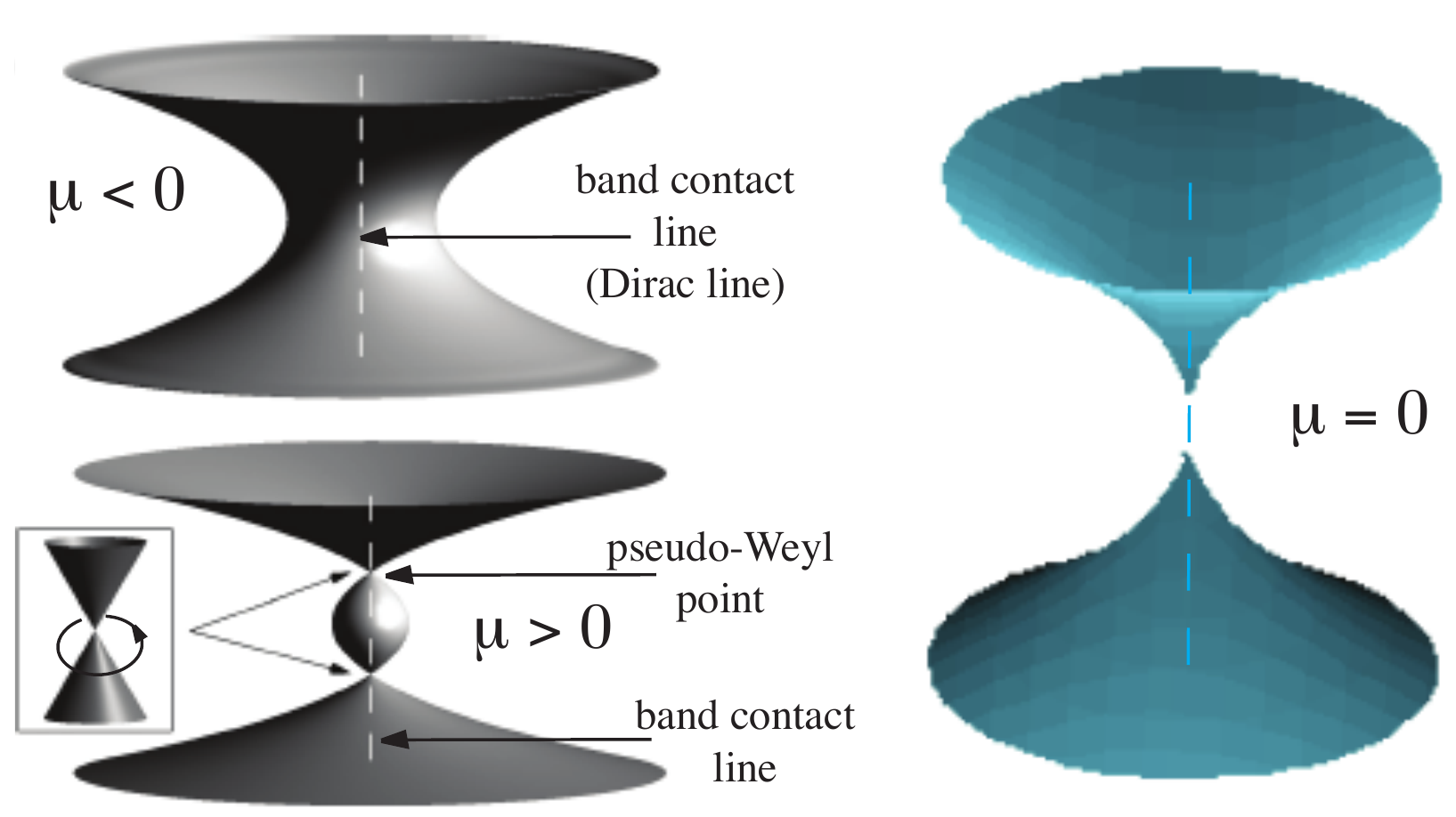}}
\medskip
\caption{Illustration of Lifshitz transition in the 3D system with Rashba spin-orbit interaction in Eq.(\ref{Rashba}) in the vicinity of the transition. The left part of figure reproduces Fig. 1 in Ref.\cite{Mikitik2014}, where the transition has been considered for the general case. ({\it left top}): The Fermi surface at $\mu<0$. It embraces the band contact line (dashed line) -- the Dirac line in the fermionic spectrum of the traceless part of the Hamiltonian (\ref{Rashba}).  ({\it left bottom}): The Fermi surface at $\mu>0$ contains two conical pseudo-Weyl points. They are situated at the Dirac line and are described by the same topological invariant, or by the invariant $N_3({\rm pseudo})$ in Eq.(\ref{NeutrinoInvariantPseudo}). ({\it inset}): The conical Fermi surface in the vicinity of a touching point. The invariant $N_2$ describing the conical Fermi surface is the integral over the shown contour.  ({\it right}): Illustration of the singular Fermi surface at the Lifshits transtion, when $\mu=0$ and the pseudo-Weyl points merge.
}
\label{RashbaMikitikFig}
\end{figure}

This transition can be illustrated using the 3D Hamiltonian with Rashba spin-orbit interaction:\cite{Rashba1959,Rashba1960}  
\begin{equation}
H= \frac{p^2}{2m} - \mu + c(\sigma_x p_y - \sigma_y p_x)\,.
\label{Rashba}
\end{equation}
Let us consider $\mu$ close to zero, i.e. $|\mu|\ll mc^2$. For $\mu<0$, the Fermi surface near the point ${\bf p}=0$ has the form in Fig. \ref{RashbaMikitikFig} ({\it left top}).
For $\mu>0$ the new Fermi pocket appears in Fig. \ref{RashbaMikitikFig} ({\it left bottom}). It is connected with the original Fermi surface by two touching points at $p_z=\pm \sqrt{2m\mu}$. The conical structure of the Fermi surface near each touching point is in Fig. \ref{RashbaMikitikFig} ({\it inset}). The conical points are situated on the Dirac lines (band contact lines \cite{Mikitik2014})  
and are characterized by the same topological invariant $N_2$.

In principle one can characterize these conical points as pseudo-Weyl points. The Hamiltonian (\ref{Rashba}) can be rewritten in the more general form 
\begin{equation}
H= m_3+ \sigma_x m_x + \sigma_y m_y\,,
\label{Rashba2}
\end{equation}
and for the vector field ${\bf m}({\bf p})$ one can write the invariant similar to that in Eq.(\ref{NeutrinoInvariant}) for the Weyl points:
\begin{equation}
N_3({\rm pseudo})=\frac{1}{8\pi}e_{ijk}\int_{S_2} dS^k ~\hat{\bf
m}\cdot \left(\frac{\partial \hat{\bf m}}{\partial {p_i}} \times \frac{\partial
\hat{\bf m}}{\partial {p_j}}\right)~.
\label{NeutrinoInvariantPseudo}
\end{equation}
The topological charges of the two pseudo-Weyl points are
$N_3({\rm pseudo})=\pm 1$.
At the Lifshitz transition, at $\mu=0$, the conical points merge, and the Fermi surface has singularity at ${\bf p}=0$ in the form
$p_z=\pm \sqrt{2mc|{\bf p}_\perp|}$ in  Fig. \ref{RashbaMikitikFig} ({\it right}).

The connection between the conical Fermi surfaces and the Weyl points can be seen if we extend the Hamiltonian (\ref{Rashba}) to the imaginary values of momenta $p_x$ and $p_y$ and then multiply it by $\sigma_z$:
\begin{equation}
\tilde H= \sigma_z H(p_x \rightarrow ip_x, p_y \rightarrow ip_y)=\sigma_z \left(\frac{p_z^2-p_x^2-p_y^2}{2m} - \mu \right)+ c(\sigma_x p_x + \sigma_y p_y)\,.
\label{Rashba3}
\end{equation}
The resulting Hamiltonian (\ref{Rashba3}) is similar to that discussed in Sec.\ref{GappedToWeylSec}, and 
 for $\mu >0$ it contains the topologically protected Weyl points at ${\bf p}=(0,0,\pm \sqrt{2m\mu})$.

 \section{Lifshitz transition with change of the shape of Dirac line}

\begin{figure}
 \includegraphics[width=0.6\textwidth]{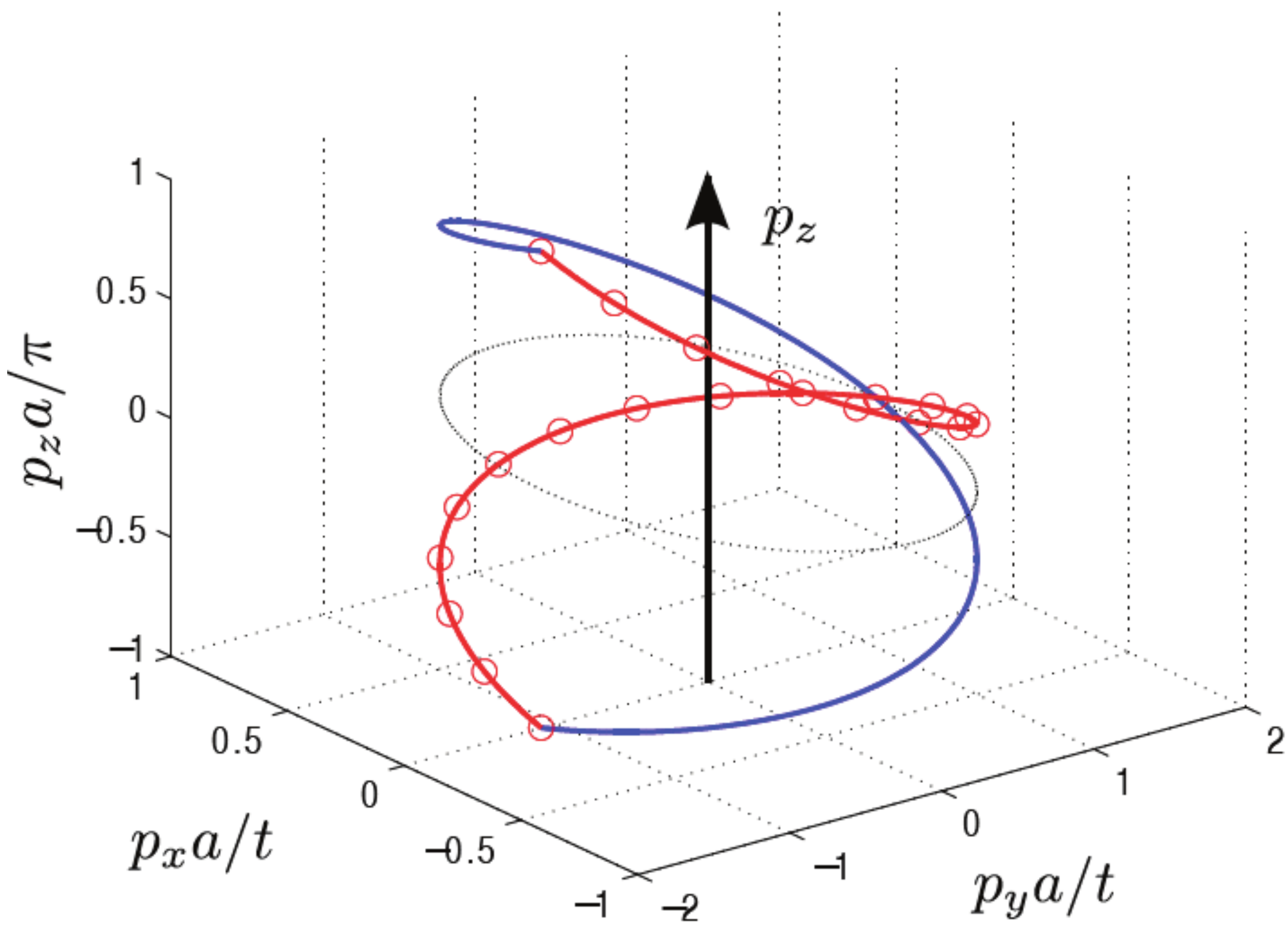}
 \caption{In rhombohedral graphite the Dirac line has the shape of a spiral  (from Ref. \cite{HeikkilaVolovik2011}). The spiral is characterized by helicity, which may change in the Lifshitz transition. Red and blue lines show the nodal lines with opposite helicity.
 At the Lifshitz transition between the two lines with opposite helicities, the nodal line becomes flat: its projection on the plane $p_z=0$  shrinks to the line segment with
zero area. According to the bulk-boundary correspondence, the projection of the nodal line to the surface determines the boundary of the $2D$ surface flat band. At the Lifshitz transition, when the Dirac line is flat, the area of the flat band shrinks to zero. 
}
 \label{spiralFig}
\end{figure}

As distinct from the Bernal graphite, in the rhombohedral graphite, the nodal line with $N_2=1$ has the shape of 
the spiral in Fig. \ref{spiralFig}.\cite{HeikkilaVolovik2011,HeikkilaKopninVolovik2011,Spiral2013} 
The peculiar Lifshitz transition in such system is related to the change of the shape of the Dirac line:
this is the transtion at which
the helicity of the Dirac line changes to the opposite. Exactly at the transition the projection of the nodal line
to the $p_z = 0$ plane shrinks to the line segment with
zero area. Simultaneously, due to the bulk-boundary correspondence, the flat band on the surface of the system is transformed
to the Fermi line.\cite{HeikkilaVolovik2011}

\section{Dirac line at Lifshitz transition between type-I and type-II Weyl points}

Let us consider again the $p$-wave superfluid/superconductor, but now in the presence 
of the superflow with velocity ${\bf v}=v\hat{\bf x}$ along the $x$ axis. The Hamiltonian (\ref{Pwave}) becomes the
 Doppler shifted:
\begin{equation}
H=p_x v + \tau_3 \epsilon_{\bf p} + c(p_x \tau_1 + p_y\tau_2) \,\, \,\,, \,\,  \,\, \epsilon_{\bf p}  = \frac{p_x^2 + p_y^2 +p_z^2}{2m}-\mu   
\,.
\label{PwaveFlow}
\end{equation}

\begin{figure}
 \includegraphics[width=0.8\textwidth]{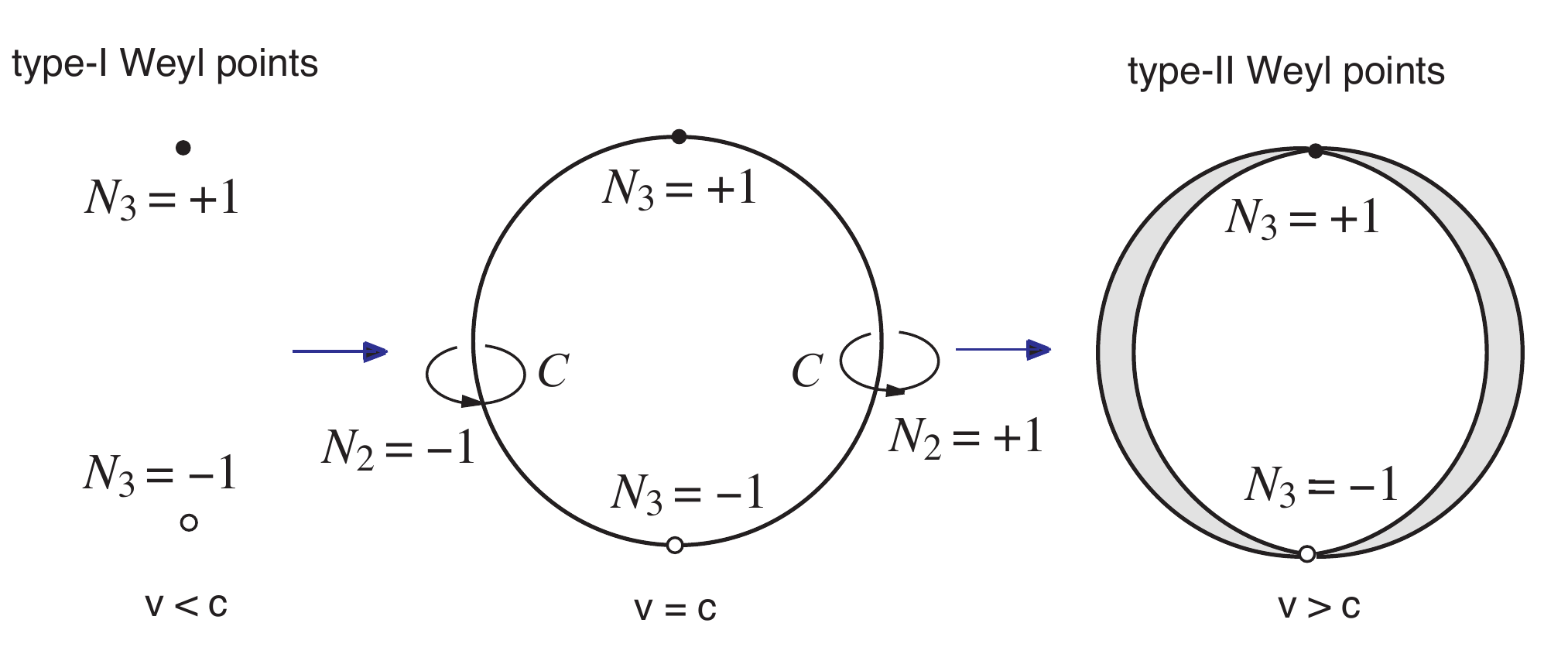}
 \caption{Lifshitz transition with formation of type-II Weyl points in chiral superfluid/superconductor in the presence of supercurrent with velocity ${\bf v}=v\hat{\bf x}$. 
At $v<c$ there are two Weyl points in spectrum of $^3$He-A . When the flow velocity $v$ exceeds the "speed of light" (the pair-breaking velocity) $c$), the Weyl cones are overtilted and form two pockets of Fermi surfaces, which touch each other at the Weyl points. In these regime the Weyl points are called the type-II Weyl.\cite{Soluyanov2015}    At the Lifshitz transition between the ordinary type-I Weyl points to the type-II Weyl ponits, which occurs at the critical velocity $v=c$, there are Dirac lines, which connect the Weyl points.
}
 \label{ChiralFlow}
\end{figure}

\begin{figure}
 \includegraphics[width=0.4\textwidth]{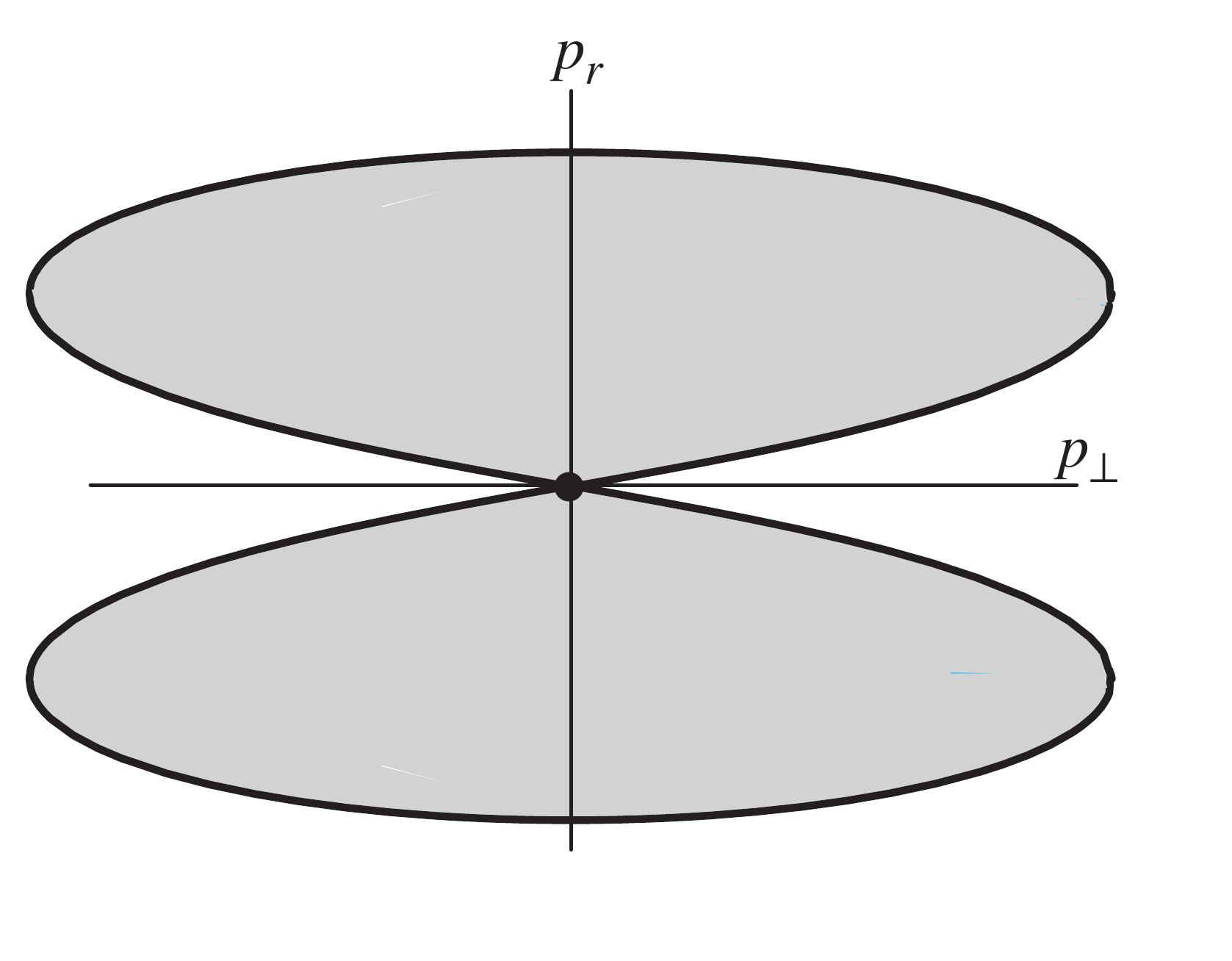}
 \caption{Type-II Weyl point and the corresponding Fermi surfaces behind the black hole horizon, from Refs.\cite{HuhtalaVolovik2002,Volovik2003}. The velocity ${\bf v}$ here is the velocity of the free falling observer, 
${\bf v}(r)= -c \hat{\bf r}\sqrt{r_g/r}$, where $r_g$ is the radius of the horizon.
Inside the horizon, where $|{\bf v}|>c$, the cone is overtilted and Weyl or Dirac  fermions transform to the type-II Weyl or type-II Dirac fermions. The Lifshitz transition between the type-I and type-II fermions takes place at the horizon, where $|{\bf v}|=c$.
}
 \label{BH}
\end{figure}

For $v<c$ the spectrum as before contains two Weyl points at ${\bf p}_\pm=(0,0,\pm \sqrt{2m\mu})$, which have opposite topological charges $N_3=\pm 1$, see Fig. \ref{ChiralFlow} ({\it left}). As distinct from the static situation at $v=0$, the Weyl cones are tilted at $v>0$. At $v=c$ the Lifshitz transition takes place 
between the ordinary (type-I) Weyl points at $v<c$ to the type-II Weyl points at $v>c$, which are the touching points of the Fermi surface pockets. Exactly at the Lifshitz transition, at $v=c$, there are two lines described by the invariant $N_2=1$.

The type-II Weyl fermions may appear in relativistic theories too,
for example behind the black hole event horizon in Fig. \ref{BH}.\cite{HuhtalaVolovik2002,Volovik2003}  
They also emerge if the relativistic Weyl fermions are not fundamental, but belong to the low energy sector of the fermionic quantum vacuum.\cite{VolovikZubkov2014}
The Hamiltonian describing the Weyl fermions in the vicinity of the topologically protected Weyl point at ${\bf p}^{(0)}$ has the general form
\begin{equation}
H= e_k^j(p_j-p^{(0)}_j) \hat\sigma^k  + e_0^j(p_j-p^{(0)}_j)\,.
\label{HamiltonianGeneral} 
\end{equation}
This expansion suggests that position ${\bf p}^{(0)}$ of the Weyl point plays the role of the vector potential of the effective $U(1)$ gauge field, while $e_k^j$
and $e_0^j$ play the role of the emergent tetrad fields.
The energy spectrum of the Weyl fermions depends on the ratio between the two terms in the rhs of Eq.(\ref{HamiltonianGeneral} ), i.e. on the parameter 
$|e^j_0 [e^{-1}]_j^k| $.\cite{VolovikZubkov2014} When
$|e^j_0 [e^{-1}]_j^k|  < 1$ one has the conventional Weyl point with the tilted Weyl cone. At
 $|e^j_0 [e^{-1}]_j^k| > 1$  two Fermi surfaces appear, which touch each other at the Weyl point. 

For a simple choice of the tetrad field, $e_k^j=c\delta_k^j$ and $e_0^j=-v\hat x^j$, where $v$ is parameter,
corresponding to the superfluid velocity in chiral superfluids, one has:
 \begin{equation}
 H=  
c{\mbox{\boldmath$\tau$}} \cdot{\bf p}  +vp_x \,.
 \label{HamiltonianSimple}
 \end{equation}
For $v>c$ the type-II Weyl point is formed. At Lifshitz transition, at $v=c$, there is the Dirac line on the $p_x$-axis with the topological charge $N_2=1$.

 Correspondingly the black hole horizon can be simulated in Weyl semimetals by creation of the interface separating the regions with type-I and type-II Weyl or Dirac points. \cite{Volovik2016} In equilibrium such effective horizon is not radiating, but the process of equilibration by filling of the electronic states inside the "black hole" will be accompanied by Hawking radiation. The Hawking temperature in the Weyl semimetals can reach the room temperature, if the black hole region is sufficiently small, and thus the effective gravity at the horizon is large.

The $2D$ version of the tilted Dirac cone has been considered in Ref. \cite{Goerbig2008}.

 \section{Lifshitz transition at which Fermi surface changes its global topological charge $N_3$}

Let us consider the relativistic model dscribed by the Hamiltonian:\cite{KlinkhamerVolovik2005a,Volovik2007} 
\begin{equation}
H=\tau_3\left({\mbox{\boldmath$\sigma$}} \cdot({\bf p} -\tau_3{\bf A})-\mu\right) + M\tau_1
\,.
\label{KissH}
\end{equation}
Here the speed of light $c=1$; ${\bf A}$ is the chiral vector potential; $\mu$ is the chiral chemical potential; and $M$ the mass. 

\begin{figure}
\centerline{\includegraphics[width=0.9\linewidth]{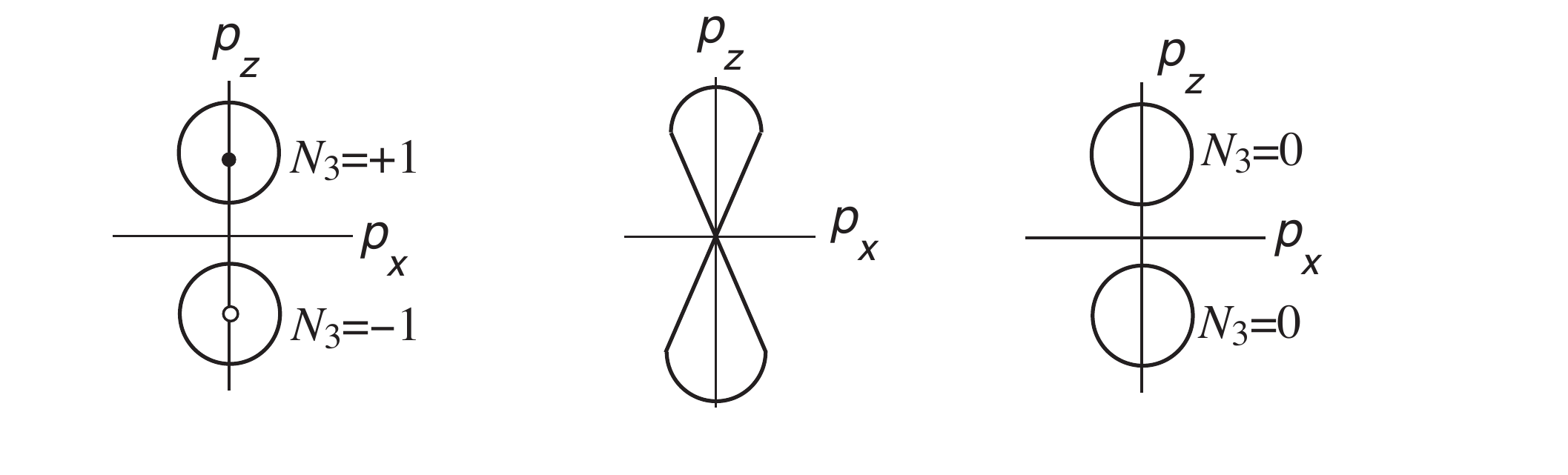}}
\medskip
\caption{Lifshitz transition with exchange of the Weyl charge (from the relativistic model discussed in Ref.
\cite{KlinkhamerVolovik2005a}). The vacua on both sides of the transition have Fermi
surfaces. On the left side of transition, Fermi surfaces contain Weyl points with nonzero
topological  charges  $N_3=+1$ and  $N_3=-1$.   At the transition, the
Fermi surfaces ``collide'', and their topological global charges $N_3$ annihilate
each other. On the right side of transition, Fermi surfaces become globally trivial, $N_3=0$.}
\label{FSTouchFig2}
\end{figure}

For $M=0$ and ${\bf A}^2>\mu^2$ one has two Weyl points with topological charges $N_3=\pm 1$, surrounded by the Fermi surfaces in Fig. \ref{FSTouchFig2} ({\it left}). It means that each Fermi surface has the global topological charge in Eq.(\ref{NeutrinoInvariant}), where the sphere $S_2$ embraces the whole Fermi sirface. With increasing mass $M$ the Weyl points move to the boundaries of Fermi surfaces. At $M^2={\bf A}^2-\mu^2$ two Fermi surfaces touch each other at ${\bf p}=0$, and  their Weyl points merge at  ${\bf p}=0$,  Fig. \ref{FSTouchFig2} ({\it middle}). This is the Lifshitz transition.
Above the transition, at  $M^2>{\bf A}^2-\mu^2$, the Fermi surfaces split again, but now they do not contain the Weyl charge, $N_3=0$,  i.e. the Fermi surfaces are globally trivial, see Fig. \ref{FSTouchFig2} ({\it right}).

This Lifshitz transition demonstrates the interplay of two topological invariants, the invariant $N_1$, which is responsible for the local stability of the Fermi surface, and the invariant $N_3$, which describes the global property of the Fermi surface.

 \section{Lifshitz transitions between gapped states}

The parameter, which crosses the point of the Lifshitz transition, such as $\mu$ in figures \ref{ParHypLifshitzFig},  \ref{FromDiracToWeyl} and \ref{FromWeylToLine}, or $v$ in figures \ref{TrigonalFig} and \ref{ChiralFlow}, provides the new dimension in addition to the momentum space parameters ${\bf p}$. This leads to new topological invariants emerging in the extended parameter space, $({\bf p},\mu)$, and as a result to numerous types of the topological  Lifshitz transitions. The correspondence between the Lifshitz transition and the singularity in the spectrum in higher dimension has been considered in Sec. \ref{QuadraticTouching}
on example of bilayer graphene and nexus point in Bernal graphite.

\begin{figure}
\centerline{\includegraphics[width=0.5\linewidth]{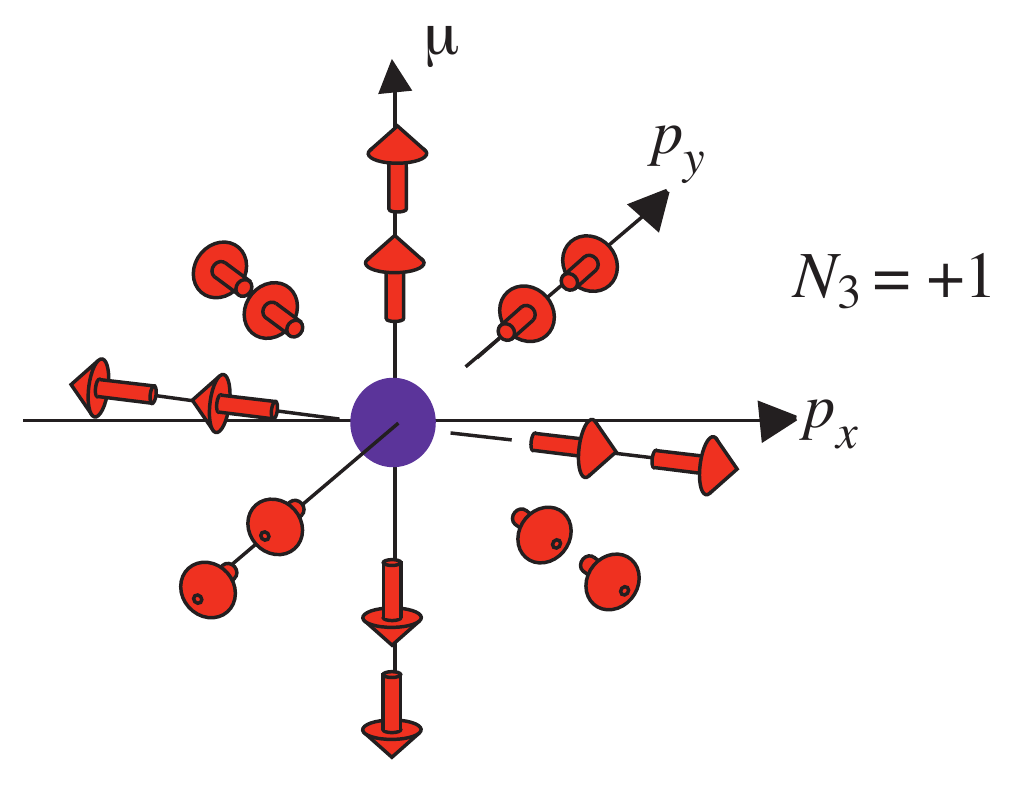}}
\medskip
\caption{The Lifshits transition between the fully gapped vacua in $2D$ system represents the topologically stable node -- the hedgehog in $(p_x,p_y,\mu)$-space. Transition occurs at $\mu=0$, where the energy spectrum has node at $p_x=p_y=0$ (center of the hedgehog). For $\mu\neq 0$ the system has gap and represents the topological insulator at $\mu>0$ and trivial insulatro at $\mu<0$.
 Compare with the Weyl point in $(p_x,p_y,p_z)$-space in Fig. \ref{FP1Fig}.}
\label{FPFig}
\end{figure}

Here we consider another example of such correspondence --  the Lifshitz transitions between the gapped states in insulators and in fully gapped superfluids/superconductors, which have the same symmetry but different values of some topological invariant.  There are many types of the topological invariants for gapped systems, which also depend on the type of the symmetry.\cite{Schnyder2008,Schnyder2009} However, typically topological classification of the gapped topological states in the $D$-dimensional material can be obtained by dimensional reduction of the classification of the gapless topological states\cite{Horava2005} in dimension  $D+1$.
That is why the Lifshitz transition between the gapped states, which occurs via the gapless states, can be described as the topological node on the spectrum in the extended $D+1$ space, which includes the $D$-dimensional momentum space and the $1D$-space of the parameter of the transition, such as $\mu$ and $v$.

For $D=2$ the instructive model Hamiltonian  is
\begin{equation}
H(p_x,p_y, \mu)= \tau_3 \epsilon_{\bf p} + c(p_x \tau_1 + p_y\tau_2) \,\, \,\,, \,\,  \,\, \epsilon_{\bf p}  = \frac{p_x^2 + p_y^2}{2m}-\mu   
\,.
\label{Pwave2D}
\end{equation}
Eq.(\ref{Pwave2D}) describes spinless fermions in a thin film of the chiral $p$-wave superfluid/superconductor. 
It is the dimensional reduction of the $3D$ Hamiltonian in Eq.(\ref{Pwave}). The superfluid is topological for $\mu>0$ and topologically trivial for $\mu<0$. At the Lifshitz transtion, which occurs at $\mu=0$, the superfluid becomes gapless. In the $2+1$ extended space $(p_x,p_y,\mu)$ the spectrum $E(p_x,p_y, \mu)$ has the point node at $(p_x,p_y,\mu)=(0,0,0)$. This node is analogous to the Weyl point in $3D$ and thus is topologically stable. It is described by the topological invariant $N_3$ in Eq.(\ref{NeutrinoInvariant}), where the integral is over the surface in the extended $2+1$ space around the node. The value of the integral is $N_3=1$.

On the other hand the integral around the node can be represented in terms of the integrals over the $(p_x,p_y)$-plane at fixed $\mu$:
\begin{equation}
N(\mu)=\frac{1}{4\pi}\int dp_xdp_y~\hat{\bf
m}\cdot \left({\partial \hat{\bf m}\over\partial {p_x}} \times {\partial
\hat{\bf m}\over\partial {p_y}}\right)~.
\label{Nonmu}
\end{equation} 
The topological invariant $N_3=1$
describing the Lifshitz transition is equal to the difference between the topological invariants describing two fully gapped systems with opposite $\mu$: 
\begin{equation}
N_3=N(\mu>0)-N(\mu<0) ~.
\label{Reduction}
\end{equation} 
In a given case one has the topological insulator at $\mu >0$ with $N(\mu>0)=1$  and the topologically trivial one  at $\mu <0$ with $N(\mu<0)=0$.

 \section{Lifshitz transition with crossing zero of Green's function}

When the interaction between electrons is important, one should use the Green's function instead of the Hamiltonian. The consideration of the Green's function leads to new types of the Lifshitz transition.
Let us consider for example the following Green's function, which may emerge in the interacting systems:
\begin{equation}
 G=\frac{1}{2} \left(\frac{1-a}{i\omega - {\bf \sigma}\cdot {\bf p}} + \frac{1+a}{i\omega+ {\bf \sigma}\cdot {\bf p}} \right)= \frac{-i\omega + a{\bf \sigma}\cdot {\bf p}}{\omega^2 + p^2} \,.
\label{ZeroGreen}
\end{equation}
The point $(\omega=0, {\bf p}=0)$ is the singular point of the Green's function, which is protected by $N_3$ topological invariant for the Weyl point, when it is written in terms of the Green's function.\cite{Volovik2003} At  $a>0$ the Weyl point has $N_3=+1$, and at $a<0$ it has $N_3=-1$. 
Lifshits topological transition, at which chirality of the fermion changes to the opposite, takes place at $a=0$ where the Green's function describes the non-chiral fermions:
\begin{equation}
 G(a=0)=\frac{1}{2} \left(\frac{1}{i\omega -p} + \frac{1}{i\omega+ p} \right) \,.
\label{ZeroGreen2}
\end{equation}
At the point of transition the Green's function crosses zero value at $\omega=0$ for all momenta ${\bf p}\neq 0$. The role of zeroes in the topology of Green's function is discussed in Refs.\cite{Volovik2007,EssinGurarie2011}

 \section{Conclusion}

\begin{figure}
\centerline{\includegraphics[width=0.7\linewidth]{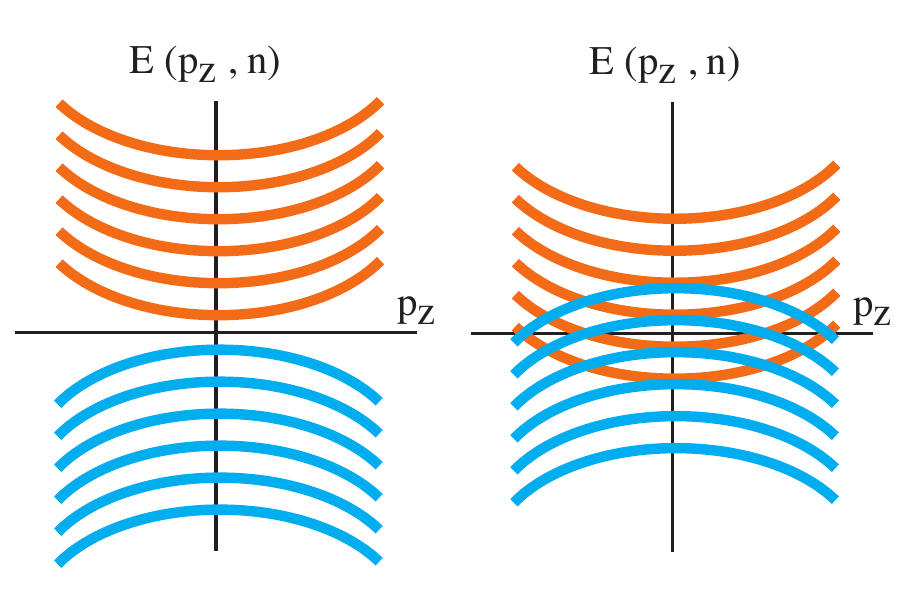}}
\medskip
\caption{Lifshits transitions in the core of vortex: formation of 1D Dirac points.\cite{MakhlinVolovik1995} 
The vortex in $s$-wave superconductor is considered, with magnetic field strong enough to shift the energy levels. Some levels cross zero energy forming the $1D$ Fermi-surfaces. Opposite spins cross zero in opposite directions, as a result the $1D$ Fermi surfaces become the $1D$ Dirac points.}
\label{VortexTrFig}
\end{figure}

Topology provides many different types of Lifshitz transition. There exists the whole Zoo of topological invariants, which describe topology in real space, in momentum space and in combined phase space. 
All of them can be changed under modification of the parameters of the system, giving rise to exotic Lifshitz transitions. One should add the change of the geometry of the shape of the manifolds of zeroes and
interconnection of the manifolds of different dimensions  (point nodes, nodal lines, nodal surfaces
and higher hypersurfaces in the phase space).

 Let us also mention the Lifshitz transitions for fermions living in the core of topological defects. Vortices in conventional $s$-wave superconductor contain the Caroli-de Gennes-Matricon levels in Fig. \ref{VortexTrFig} ({\it left}).\cite{Caroli}
 Typically these vortices do not contain the gapless modes: there is a small gap of order $\Delta^2/E_F \ll \Delta$, the so-called minigap.  However, if the effect of the Pauli magnetic field is strong enough few lowest energy levels may cross zero  energy. In such Lifshitz transition the pairs of $1D$ Dirac points are formed in Fig. \ref{VortexTrFig} ({\it right}).\cite{MakhlinVolovik1995} Contrary to the $s$-wave superconductors, vortices in superfluid $^3$He-B contain a large number  (on the order of $E_F/\Delta\sim 10^3$) one-dimensional Fermi surfaces or Dirac points.\cite{SilaevVolovik2014} By changing the magnidude of applied magnetic field one has many Lifshitz transitions.
 
Lifshitz transitions are ubiquitous. They take place not only condensed matter, but also in the relativistic vacua.
In particular, confinement-deconfinement transition at  zero temperature can be also considered as Lifshitz transition, at which the spectrum of quarks radically changes without symmetry breaking.

\section*{Acknowledgements}
I thank E.I. Rashba for interesting discussions.
This work has been supported in part  by the Academy of Finland
(project no. 250280), and by the facilities of the Cryohall
infrastructure of Aalto University.

\end{document}